\documentclass[letter]{aa}  
\usepackage{textcomp}
\usepackage{graphicx}
\usepackage{txfonts}
\begin{document} 

   \title{Yet another UFO in the X-ray spectrum of a high-z lensed QSO}

  \author{M. Dadina\inst{1}, C. Vignali\inst{1,2}, M. Cappi\inst{1}, G. Lanzuisi\inst{1,2}, G. Ponti\inst{3}, E. Torresi\inst{1,2}, B. De Marco\inst{4}, G. Chartas\inst{5}
, M. Giustini\inst{6} }

   \institute{INAF/OAS, Osservatorio di Astrofisica e Scienza dello Spazio di Bologna, via Gobetti 93/3, 40129, Bologna, Italy  \email{dadina@iasfbo.inaf.it}
                           \and
 Dipartimento di Fisica e Astronomia dell'Universit\'a degli Studi di Bologna, via P. Gobetti 93/2, 40129, Bologna, Italy
 \and
Max-Planck-Institut f{\"u}r Extraterrestrische Physik, Giessenbachstrasse 1, D-85748, Garching, Germany
\and
Nicolaus Copernicus Astronomical Center, Polish Academy of Sciences, Bartycka 18, PL-00-716 Warsaw, Poland
\and
Department of Physics and Astronomy of the College of Charleston,  Charleston, SC 29424, USA
\and
SRON Netherlands Institute for Space Research, Sorbonnelaan 2, 3584 CA Utrecht, the Netherlands}

   \date{Received  ;  Accepted}
\authorrunning{Dadina et al.}

\titlerunning{UFO detection in the X-ray spectrum of the z=2.64 QSO MG J0414+0534}

% \abstract{}{}{}{}{} 
% 5 {} token are mandatory
 
  \abstract
   {}
   { Ultra-fast outflows (UFO) appear to be common in local active galactic nuclei (AGN) and may be powerful enough ($\dot{E}_{kin}$$\geq$1\% of L$_{bol}$) to effectively quench the star formation in their host galaxies. To test feedback models based on AGN outflows, it is mandatory to investigate UFOs near the peak of AGN activity, that is, at high-z where only a few studies are available to date.}
   {UFOs produce Fe resonant absorption lines measured above $\approx$7 keV. The most critical problem in detecting such features in distant objects is the difficulty in obtaining X-ray data with sufficient signal-to-noise. We therefore selected a distant QSO that gravitational lensing made bright enough for these purposes, the z=2.64 QSO MG J0414+0534, and observed it with XMM-Newton for $\approx$78 ks.}
   {The X-ray spectrum of MG J0414+0534 is complex and shows signatures of cold absorption (N$_{H}\approx$4$\times$10$^{22}$ cm$^{-2}$) and of the presence of an iron emission line (E$\approx$6.4 keV, EW$=$95$\pm$53 eV) consistent with it originating in the cold absorber. Our main result, however, is the robust detection (more than 5$\sigma$) of an absorption line at E$_{int}\approx$9.2 keV (E$_{obs}\approx$2.5 keV observer frame). If interpreted as due to FeXXVI, it implies gas outflowing at $v_{out}\approx$0.3c. To our knowledge, this is the first detection of an UFO in a radio-loud quasar at z$\geq$1.5. We estimated that the UFO mechanical output is $\dot{E}_{kin}$$\approx$2.5$L_{bol}$ with $\dot{p}_{out}/\dot{p}_{rad}\approx$17 indicating that it is capable of installing significant feedback between the super-massive black hole (SMBH) and the bulge of the host galaxy. We argue that this also suggests a magnetic driving origin of the UFO.}
   {}


   \maketitle
%
%________________________________________________________________

\section{Introduction}

Since the discovery of the relation between the mass of a super-massive black hole (SMBH) and the bulges of their host galaxies ({\it i.e.,} the ``M$_{\bullet}-\sigma$ relation'', Kormendy \& Richstone 1995, Magorrian et al. 1998), we know that SMBHs likely play a role in the formation and growth of the galaxies (Fabian 2012). Active galactic nuclei (AGN)-driven ultra-fast outflows (UFOs) (Tombesi et al 2010a) have been recently proposed as a major feedback process whereby sweeping out and/or compressing the interstellar gas may influence the formation and growth of the galaxies (Fabian 2012, King \& Pounds 2015).  

Resonant absorption lines detected in the $\sim$7-10 keV energy range due to highly ionized iron are the UFO signatures. They are measured in both radio-quiet and radio-loud objects and both in type-I and -II AGNs (Tombesi et al. 2010a, 2010b; Gofford et al. 2013). While the average properties of UFOs are known at low z, we have only a few UFO detections at z$\geq$1.5, that is, where they may have acted to shape the M$_{\bullet}-\sigma$ relation seen today (Hasinger et al. 2002; Chartas et al. 2002a, 2003, 2007, 2016; Lanzuisi et al. 2012; Vignali et al. 2015).

Here we present the XMM-Newton spectrum of MG J0414+0534, a lensed (magnification factor $\mu\sim$30-60, Trotter, Winn, \& Hewitt, 2000; Minezaki et al. 2009) and radio-loud type-I QSO at z=2.64 (Lawrence et al. 1995). The target is also a hyper-luminous infrared and red QSO  (Lawrence et al. 1995; McLeod et al. 1998). These objects are thought to represent a dust-enshrouded phase in AGN evolution during which nuclear winds are expected to be present and expel/heat the cold gas in the hosting galaxy (Georgakakis et al. 2009; Urrutia et al. 2009) thus enabling feedback processes between the SMBH and the galaxy bulges (Fabian 2012, King \& Pounds 2015). In X-ray, the source was previously  pointed by Chandra and the spectrum was described by an absorbed power-law  ($\Gamma=$1.7$\pm$0.1, N$_{H}=$4.7$\pm$0.7$\times$10$^{22}$ cm$^{-2}$, errors are at 90\% confidence level for one parameter of interest here and throughout the paper, Avni 1976) plus an iron line in emission (E$_{FeK\alpha}=$6.4$\pm$0.1 keV, EW$_{FeK\alpha}=$200$\pm$100 eV; Chartas et al. 2002).

\section{Data reduction and analysis}

XMM-Newton pointed to MG J0414+0534 on March 11, 2017. SAS-15 and the latest available software and response matrices were used to reduce and analyze the data. The observation lasted $\approx$78 and $\approx$76 ks for EPIC-pn and EPIC-MOS instruments, respectively. Since it was affected by soft-p$^{+}$ flares, high-background intervals were removed through an iterative sigma-clipping  procedure applied to the 10-15 keV band data; we were left with cleaned exposure times of 48.5, 66.5, and 69.1 ks of exposure for pn, MOS1, and MOS2, respectively.

   \begin{figure}
   \centering
   \includegraphics[width=5.cm, height=8.4cm, angle=-90]{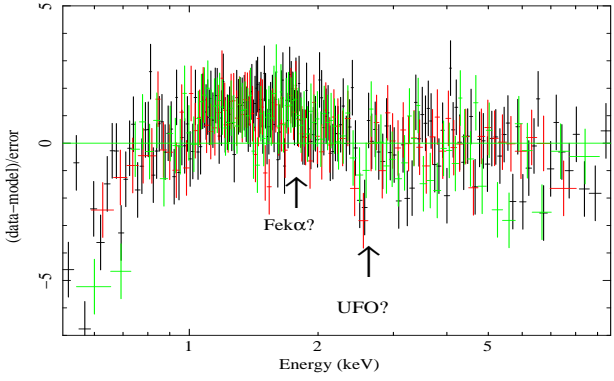}
      \caption{Data-to-model ratio expressed in terms of standard deviations with respect to a power-law absorbed by Galactic column density (1.02$\times$10$^{21}$ cm$^{-2}$, Kalberla et al. 2005). A deep and narrow drop of counts at E$\approx$2.5 keV (observed frame) is clearly present. Black data points indicate EPIC-pn, while green and red data points represent EPIC-MOS1 and MOS2, respectively.}
   \end{figure}

   \begin{figure}
   \centering
   \includegraphics[width=3.5cm, height=8.4cm, angle=-90]{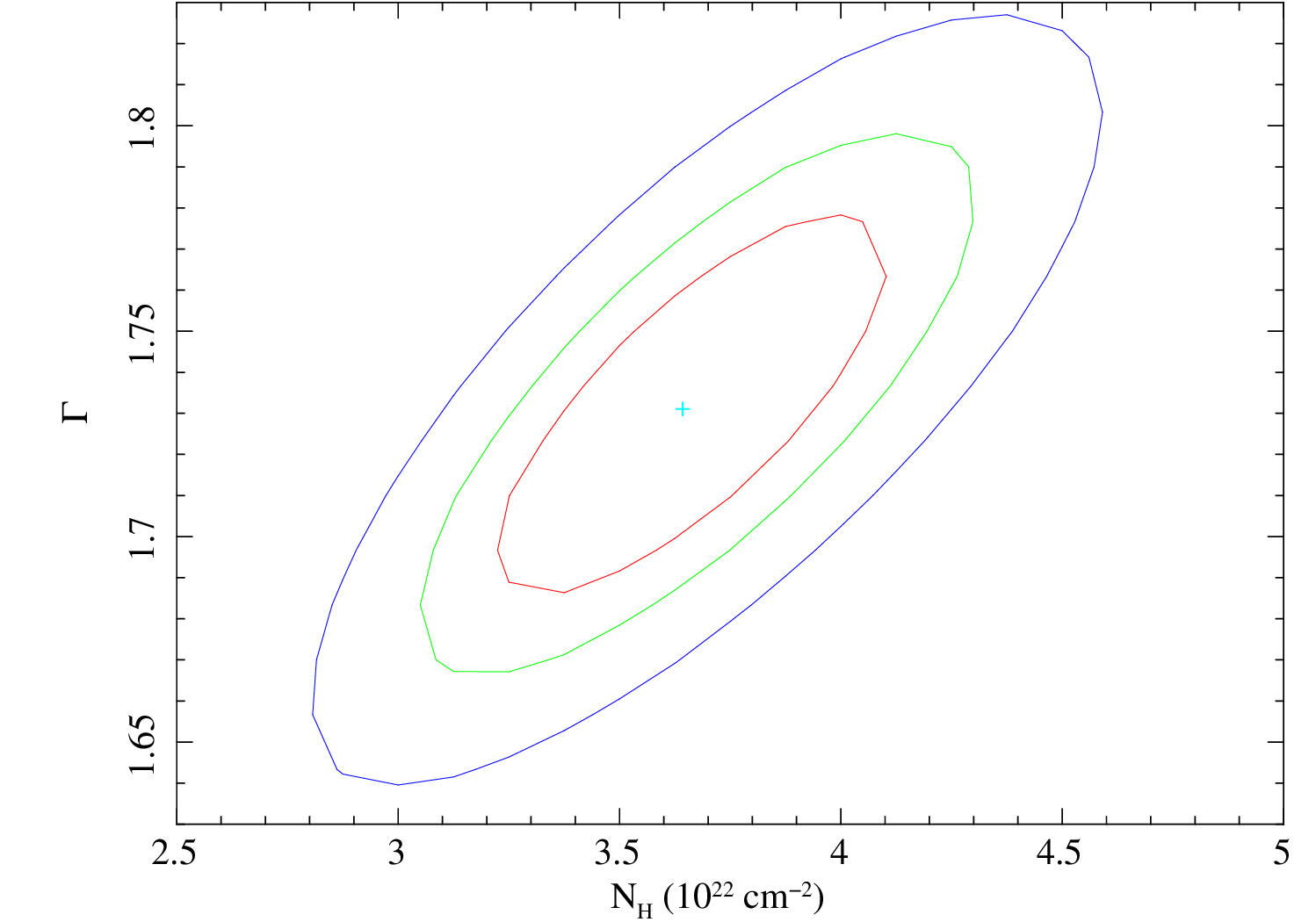}
   \includegraphics[width=3.5cm, height=8.4cm, angle=-90]{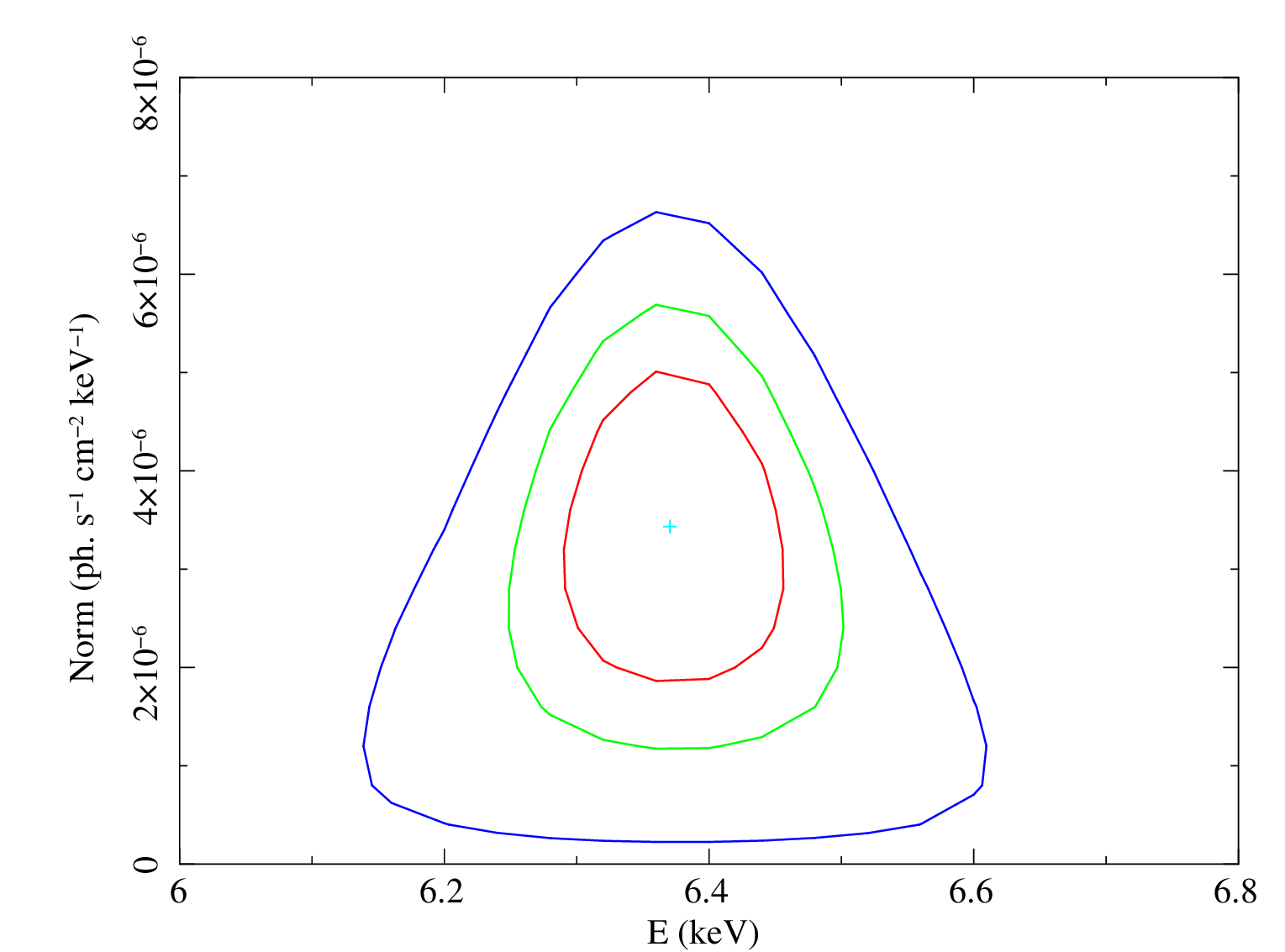}
      \caption{Confidence contours of the photon index vs. the absorbing column {(upper panel)} and{} of the Gaussian emission line normalization vs. its energy centroid (lower panel) (rest frame) adopting  model \#1 in Table 1. Contours are at 99, 90 and 68\%.}
   \end{figure}

The images of MG J0414+0534 are within $\approx$3'' (Chartas et al. 2002), thus they form a single ``point-like'' source in XMM-Newton. Source counts  ($\approx$9000 in total in the 0.3-10 keV band) have been extracted from circular regions with radii of 25''  for pn and 20'' for the MOS. Background was extracted from larger, source-free circular regions in the same chip of the target; it contributes  $\approx$5-10\% of the flux in the 2-5 keV band. 

Since no significant variability was observed, we performed a time-averaged spectral analysis. Data were grouped so as to obtain 20 source plus background counts per bin. Spectral features are clearly visible once the spectrum is modeled with a power-law and, among them, a drop of counts at E$\approx$2.5 keV (Fig. 1).  The X-ray flux of MG J04141+0534 is F$_{0.5-8 keV}\approx$4$\times$10$^{-13}$ erg s$^{-1}$  cm$^{-2}$  (see model \#1 in Table 1) which is within the range of values previously recorded (F$_{0.5-8 keV}\approx$2.5-8$\times$10$^{-13}$ erg s$^{-1}$  cm$^{-2}$, Chartas et al. 2002; Pooley et al. 2012). Following Chartas et al. (2002) we modeled the XMM-Newton spectrum with an absorbed power law plus an iron line in emission finding consistent results, that is,  $\Gamma$$=$1.75$\pm$0.05, N$_{H}=$3.9$\pm$0.5$\times$10$^{22}$ cm$^{-2}$, E$_{FeK\alpha}=$6.37$\pm$0.10 keV with an equivalent width EW$_{FeK\alpha}=$95$\pm$53 eV.

  \begin{table*}
\tiny
      \caption[]{Spectral models. {\bf Upper table}  Column 1: Model number; Column 2: absorbing column in excess to the Galactic value; Column 3: photon index; Column 4: energy of the emission line; Column 5: emission line rest frame EW; Column 6: energy of the absorption line;  Column 7: absorption line EW; Column 8: 0.5-8 keV flux; Column 9: 2-10 keV flux; Column 10:  $\chi^2$/d.o.f.
{\bf Lower table}  Columns 1-5 as in upper table. Column 6: column density of the ionized absorber; Column 7: Log of the ionization parameter expressed in  erg s$^{-1}$ cm; Column 8: observed redshift of the ionized absorber; Column 10:  $\chi^2$/d.o.f. Line widths are fixed to 0 eV.}
         \label{tab1}
         \centering
         \begin{tabular}{lcccccccccl}
 \hline\hline
  &&&&&&&&&\\
 \# & N$_{H}$          & $\Gamma$ & E$_{FeK\alpha}$ & EW$_{FeK\alpha}$& Eabs & EWabs & F$_{0.5-8 keV}$                  &F$_{2-10 keV}$                & $\chi^2$/d.o.f.\\
  &&&&&&&&&\\
  & 10$^{22} cm^{-2}$ &          & keV           &  eV          & keV  &  eV   &   10$^{-13}$ erg s$^{-1}$ cm$^{-2}$&10$^{-13}$ erg s$^{-1}$ cm$^{-2}$&               \\ 
  &&&&&&&&&\\
\hline
 &&&&&&&&&\\
1  & 3.86$^{+0.47}_{-0.50}$&1.75$^{+0.05}_{-0.05}$&6.37$^{+0.10}_{-0.10}$&105$^{+54}_{-51}$&&&3.82&3.29&333.5/349\\
 &&&&&&&&&\\
2 & 3.71$^{+0.60}_{-0.50}$&1.73$^{+0.05}_{-0.05}$&6.37$^{+0.11}_{-0.11}$&94$^{+53}_{-52}$&9.22$^{+0.11}_{-0.10}$&235$^{+74}_{-75}$&&&306.4/347&\\
  &&&&&&&&&\\
\hline\hline
\end{tabular}
\raggedright
%\begin{tabular}{lcccccccccl}
%  &&&&&&&&&\\
%   &   N$_{H}$            & $\Gamma$            & r         &  $\Theta^{\circ}$ & Eabs             & EWabs      &\hspace{3.cm} & \hspace{2.5cm}  & $\chi^2$/d.o.f.\\
%  &&&&&&&&&\\
%  &10$^{21} cm^{-2}$&&&degrees&keV &eV &&\\
%  &&&&&&&&&\\
%\hline
%  &&&&&&&&&\\
%3  &4.37$^{+0.62}_{-0.60}$   & 1.90$^{+0.14}_{-0.12}$ &  $\leq$0.1 &  30$^f$         &9.22$^{+0.11}_{-0.11}$&212$^{+76}_{-78}$&&&309.4/348\\
%  &&&&&&&&&\\
%\hline\hline
%\end{tabular}
\begin{tabular}{lcccccccccl}
 &&&&&&&&&\\
  & N$_{H}$                & $\Gamma$            & E$_{FeK\alpha}$ & EW$_{FeK\alpha}$&  N$_{H, ion}$           &  Log($\xi$)                 &   z          &\hspace{6.5cm}&  \hspace{3.3cm}    & $\chi^2$/d.o.f.\\
 &&&&&&&&&\\
  & 10$^{21} cm^{-2}$       &                     & keV           &  eV          &        10$^{22} cm^{-2}$            &                           &                 &&   &               \\ 
 &&&&&&&&&\\
\hline
 &&&&&&&&&\\
3  & 3.67$^{+0.23}_{-0.22}$  & 1.69$^{+0.01}_{-0.07}$  & 6.37$^{+0.10}_{-0.11}$         & 79$^{+51}_{-54}$          &   83$^{+17,pegged}_{-50}$ &   3.89$^{+0.27}_{-0.54}$      & 1.72$^{+0.17}_{-0.17}$ && &  297.9/344 \\
  &&&&&&&&&\\
  \hline\hline

\end{tabular}

\normalsize

\end{table*}

   \begin{figure}
   \centering
   \includegraphics[width=3.5cm, height=8.3cm, angle=-90]{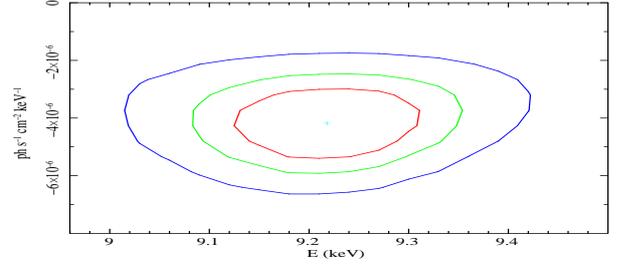}
      \caption{Confidence contours plot of the Gaussian absorption line normalization vs. its rest-frame energy centroid (see model \#2 in table 1). Contours confidence levels are as in Fig. 2.}
         \label{}
   \end{figure}

The addition of a Gaussian in absorption at E$\approx$2.5 keV (E$\approx$9.2 keV rest frame) is required by the data ($\Delta\chi^{2}$$\approx$27 for two parameters of interest corresponding, using the F-test, to a 5$\sigma$ detection; model \#2 in Table 1 and Fig. 3). Its EW is $\approx$235$\pm$70 eV (rest frame) and it is consistent to be narrow (if the line width is left free to vary we obtain $\Delta\chi^{2}\approx$0.1 and a 90\% confidence upper limit of $\sigma\leq$250 eV, rest frame). This makes an edge
origin implausible for at least part of the feature as proposed for APM 08279+5255 (Hagino et al. 2017): indeed if we substitute the Gaussian with an edge we obtain a worst fit by $\Delta\chi^{2}\approx$5 for the same number of parameters. To test the absorption feature, we searched for its presence in each single EPIC detector dataset. We used the model \#1 in Table 1 as a baseline. All the parameters of the model (except for the width of the lines which was fixed to $\sigma$=0) were free to vary. The result is plotted in Fig. 4. The absorption line is detected at more than 99\% confidence level in both MOS1 ($\Delta\chi^{2}$=14.2 for two parameters of interest) and pn datasets ($\Delta\chi^{2}$=13.0 for two parameters of interest), while there are hints of its presence in the same energy range in the MOS2 ($\Delta\chi^{2}$=3.5 for two parameters of interest). A similar combination of independent detections is highly improbable. We performed 1000 Monte-Carlo simulations for each EPIC detector using  model \#1 of Table 1 as baseline.  We searched for detections of spurious absorption lines between (rest frame) 7 and 14 keV (corresponding to outflow velocities of $\sim$0.01-0.6c) in the simulated spectra. We found that none of the 1000 simulations allowed us to obtain detections for which the line centroids are within 1 keV range for the three detectors (rest frame, see Fig. 4) and with a $\Delta\chi^2$ of at least 10 for two instruments and 3 for the other. Thus, considering the conservative approach that we used, we can assess that the probability of measuring an absorption feature as seen in  MG J0414+0534 by chance is well below 0.1\% and fully consistent with the combined probability obtained with the F-test (see above).  Since the line is close to some instrumental edges (E$\approx$2.35 and E$\approx$2.8 keV), we also tried, without success, to account for the E$\approx$2.5 keV feature allowing the detector gain to change (using "gain fit" within Xspec). We finally searched for a similar line in the longest Chandra exposure, finding that, fixing the line at E=9.2 keV, the EW is $\leq$130 eV (90\% confidence level); that is, if present, the line has varied since then (EW=235$\pm$70 eV today).

   \begin{figure}
   \centering
   \includegraphics[width=3.5cm, height=8.4cm, angle=-90]{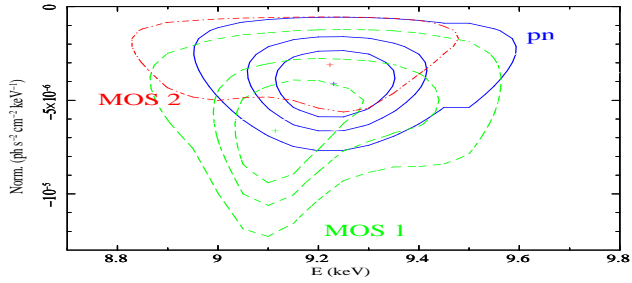}
      \caption{Confidence contours plot of the Gaussian absorption line normalization vs. its rest-rame energy for each single EPIC instrument. Contours confidence levels are as in Fig. 2.}
         \label{FigVibStab}
   \end{figure}

The detection of the FeK$\alpha$ emission line may indicate the presence of a reflection component. This feature is commonly observed in nearby Seyfert galaxies (e.g., Perola et al. 2002), and recently it has been detected also in some
high-z QSO (Dadina et al. 2016, Lanzuisi et al. 2016). To test this hypothesis and to further probe the robustness of the detection of the absorption feature against a more complex underlying continuum, we tried the Pexmon reflection model (Nandra et al. 2007) fixing the inclination angle ($\Theta$=30$^{\circ}$) and the high-energy cut-off (E$_{cut-off}$=100 keV). The data do not require this component and we obtained an upper limit (90\% confidence level) on the relative reflected-to-direct emission normalization of {r}$\leq$0.1. The detection of the absorption line is highly significant (more than 99\%) also in this case.

   \begin{figure}
   \centering
   \includegraphics[width=9.5cm, height=8.4cm, angle=-90]{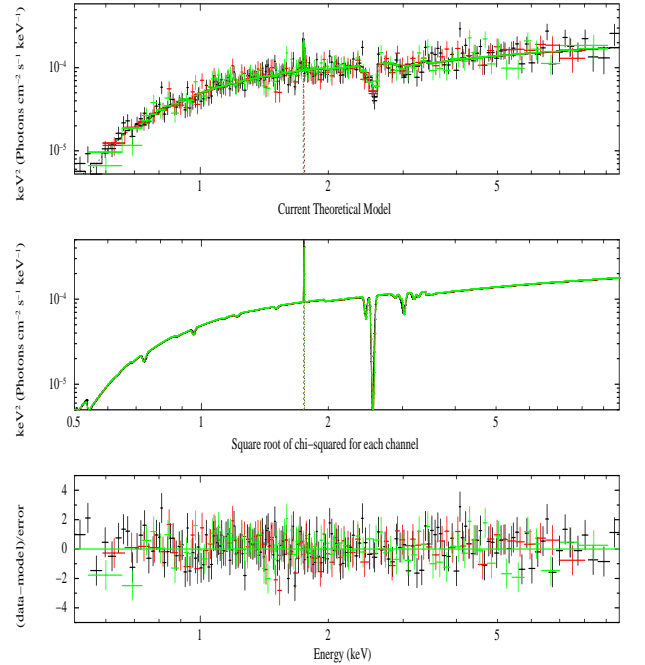}
      \caption{Unfolded X-ray observed frame energy spectrum of MG J0414+0534 ({\it upper panel}) obtained using model \# 3 in Table 1 and displayed in the {\it middle panel}. This model fits well the data
and no strong residuals are left ({\it lower panel}). Color-code is as in Fig. 1.}
   \end{figure}

We finally tried a UFO scenario in which the absorption feature at 2.5 keV is due to ionized and outflowing gas. To this end, we used the $warmabs$ model based on $Xstar$ (Kallman \& Bautista 2001). We fixed the abundances af all elements to the solar value and the turbulence velocity to {\it v$_{turb}$}=3000 km/s, in agreement with what  is measured in local AGN (Tombesi et al. 2012, 2014). The free parameters of the fit are the column of the ionized absorbing gas, its ionization parameter, and the redshift at which the absorber is detected. This last value allows us to infer the outflow velocity of the absorber. As presented in Table 1 (model \#3) and plotted in Fig. 5, we obtained a good fit and the absorber is found to be at redshift z$_{obs}\approx$1.72, which corresponds, considering the relativistic effects along the line of sight, to an outflow velocity of {\it v$_{out}=$}(0.28$\pm$0.05)c. The ionization state ($Log(\xi)\approx4$) strongly indicates that the absorption line is due to FeXXVI (see also Tombesi et al. 2011).

\section{Discussion and results}

We present the results obtained analyzing the XMM-Newton data of the radio-loud quasar MG J0414+0534 taken on March 11, 2017. We probed  its radio-loudness using the parameter R=f$_{5 GHz}$/f$_{4400\AA}$ (R$\geq$10 for radio-loud sources, Kellermann et al. 1989). To obtain the rest frame value of R we used the observed fluxes in H band (m$_H$$\approx$13.95) (Skrutskie et al. 2006) and at 1.4 GHz (f$_{1.4 GHz}$=2.1$\pm$0.1 Jy, Condon et al. 1998). The result is that R$\approx$780. The average shape of the X-ray continuum is very much in agreement with what was previously found by Chartas et al. (2002) for the brightest image (Image A) of the source. The photon index is $\Gamma\approx$1.7 and there is a cold absorbing column of N$_{H}\approx$4$\times$10$^{22}$ cm$^{-2}$. We also detected a cold iron line (E$_{FeK\alpha}=$6.4$\pm$0.1 keV) in emission with EW$_{FeK\alpha}=$95$\pm$53 eV. According to the present analysis, the iron emission line may be due to the same matter responsible for the cold absorption assuming an almost spherical distribution of such a component (e.g., Leahy \& Creighton 1993).

The observed luminosity of MG J0414+534, once corrected only for absorption, is L$_{2-10 keV}\approx$1.5$\times$10$^{46}$ erg s$^{-1}$ adopting a standard $\Lambda$CDM cosmology with H$_0$=70 km s$^{-1}$ Mpc$^{-1}$ and $\Omega_{\lambda}$=0.73. If we assume the magnification factor $\mu=$45 between the estimated values of 30 and 60 (Trotter, Winn, \& Hewitt 2000; Minezaki et al. 2009), we can infer an intrinsic X-ray luminosity L$_{2-10 keV}\approx$3$\times$10$^{44}$ erg s$^{-1}$ that corresponds to an intrinsic L$_{bol}\approx$1$\times$10$^{46}$ erg s$^{-1}$ assuming the bolometric correction factor ($k_{bol}\approx$30) by Lusso et al. (2012). Based on the H${\beta}$ broadening, the SMBH mass has been estimated to be M$_{\bullet}\approx$1.8$\times$10$^{9}$$M_{\odot}$ (Peng et al. 2006) and this implies that the source is emitting at $\approx$5\% of its Eddington limit (L$_{Edd}\approx$2$\times$10$^{47}$ erg s$^{-1}$).  

The main result of our analysis is the first detection, to our knowledge, of an UFO in a radio-loud object at  z$\geq$1.5. The absorption feature is due to iron resonant absorption (essentially FeXXVI) in ionized and outflowing gas (Log($\xi$)$\approx$3.9, N$_{H}\approx$8$\times$10$^{23}$ cm$^{-2}$) with a velocity of {\it v$_{out}$=}(0.28$\pm$0.05)c.  It is worth noting here that the UFO characteristics of MG J0414+0534 are consistent with what observed in nearby radio-loud AGNs (v$_{out}\sim$0.04-0.43c, N$_{H}\geq$10$^{22}$  cm$^{-2}$, Log($\xi$)$\sim$=1.4-5.6, Tombesi et al. 2014).  Assuming standard "recipes" and considering the large uncertainities on parameters such  as M$_{\bullet}$, $\mu$ and k$_{bol}$, we can try to infer a very rough and purely indicative estimate of the outflow mechanical output. Following Tombesi et al. (2016), we assume that the outflowing gas has been detected at a radius at  which  the  observed  velocity  corresponds to the escape velocity,  that is, $r=\frac{2GM_{\bullet}}{v^{2}_{out}}$. Using M$_{\bullet}$ and $v_{out}$ reported above, we obtain that r$\approx$6.8$\times$10$^{15}$ cm, that is, r$\approx$25r$_g$. Following Crenshaw et al. (2003), we can estimate the mass-ouflow rate as:

\begin{equation} 
 \dot{M}_{out}=4 \pi m_{p}\mu N_{H}v_{out}rC
,\end{equation}

where m$_p$ is the proton mass, $\mu$ the mean atomic mass per proton (1.4 for solar abundances), N$_{H}$ the column of the ionized gas, v$_{out}$  the line of sight outflow velocity, $r$ the absorber’s radial location, and C the global covering factor (C$\approx$0.5 here, i.e., similar at the median value obtained by Tombesi et al. 2010). The obtained mass-outflow rate is of the order of $\approx$11M$_{\odot}$ year$^{-1}$ that corresponds to $\dot{E}$$_{kin}$$\approx$2.5$\times$10$^{46}$ erg s$^{-1}$, that is, $\approx$2.5$L_{bol}$, and to an outflow momentum rate of $\dot{p}_{out}\approx6\times$10$^{36}$ g cm s$^{-2}$, that is approximately 17 times the radiation force $\dot{p}_{rad}$=$L_{bol}/c$. The $\dot{E}/L_{bol}\approx$2.5 ratio is much larger than what is observed in the local Universe ($\dot{E}$$_{kin}/L_{bol}\approx$0.01, Tombesi et al. 2013, Gofford et al. 2015) and well above the limit to switch-on/off feedback mechanisms by AGN-driven winds ($\dot{E}_{kin}/L_{bol}\geq$0.01, Di Matteo et al. 2005; Hopkins \& Elvis 2010) but it is consistent with what is observed  in other distant quasars (see for example the case of the radio-quiet quasar HS 0810+2554 where $\dot{E}$$_{kin}\approx 9L_{bol}$, Chartas et al. 2016). Moreover, together with the large ratio between the wind and radiation forces ($\dot{p}_{out}/\dot{p}_{rad}\approx$17), this may indicate that the magnetic field is probably dominating in driving the outflow in accordance with the radio loudness of the source. 

Finally, it is worth noting here that MG J0414+0534 is the seventh QSO at z$\geq$1.5 in which UFOs have been detected. Excluding HS 1700+6416 (Lanzuisi et al. 2012) and PID352 (Vignali et al. 2015) which are non-lensed, the remaining objects, namely APM 08279+5255, PG1115+080, H1413+117,  HS 0810+2554 (Hasinger et al. 2002, Chartas et al. 2003, 2007, 2009, 2016) and MG J0414+0534 are lensed. The flux enhancement due to the lensing does certainly help to collect good quality X-ray spectra and this may help in detecting such features. Alternatively, one can speculate that the flux enhancement makes it easier to probe weaker fluxes and, if the anti-correlation between the absorption line EW and the source flux observed in IRAS 13224$-$3809 (Parker et al. 2017) holds also at high-z, gravitational lensing helps in getting stronger features. However, the current absence of large enough samples of good-quality X-ray spectra of either lensed or non-lensed high-z QSO has hampered the study of these or other possible effects which must be accounted for if we want to understand how the feedback mechanism worked along cosmic time to shape the observed  M$_{\bullet}-\sigma$ relation. 
   
\begin{acknowledgements}

We thank the referee, James Reeves, for his helpful comments. This work is based on observations obtained with XMM-Newton, an ESA science mission with instruments and contributions directly funded by ESA Member States and the USA (NASA). MD acknowledges financial under contract  ASI - INAF grant I/037/12/0. GP acknowledges the Bundesministerium f\"ur Wirtschaft und Technologie/Deutsches Zentrum f\"ur Luft- und Raumfahrt (BMWI/DLR, FKZ 50 OR 1408) and the Max Planck Society. BDM acknowledges support from the European Union's Horizon 2020 research and innovation programme under the Marie Sk{\l}odowska-Curie grant agreement No. 665778 via the Polish National Science Center grant Polonez 
UMO-2016/21/P/ST9/04025.

\end{acknowledgements}

%-------------------------------------------------------------------

{}

\end{document}